\documentclass[conference]{IEEEtran}

\usepackage{cite}
\usepackage{amsmath,amssymb,amsfonts}
\usepackage{graphicx}
\usepackage{textcomp}
\usepackage{xcolor}
\usepackage[nolist]{acronym}
\usepackage{svg}
\usepackage{todonotes}
\usepackage{paralist}
\usepackage{booktabs}
\usepackage{xspace}
\usepackage{url} 
\usepackage{xcolor} 
\definecolor{greenwater}{rgb}{0.1, 0.3, 0.4}

\def\BibTeX{{\rm B\kern-.05em{\sc i\kern-.025em b}\kern-.08em
    T\kern-.1667em\lower.7ex\hbox{E}\kern-.125emX}}
\newcommand{\mypar}[1]{\smallskip\noindent\textbf{#1.}\xspace}

\begin{document}

\begin{acronym}
    \acro{ECU}[ECU]{Electronic Control Unit}
    \acro{CAN}[CAN]{Controller Area Network}
    \acro{IDS}[IDS]{Intrusion Detection System}
    \acro{LSTM}[LSTM]{Long Short-Term Memory}
    \acro{V2V}[V2V]{vehicle-to-vehicle}
    \acro{V2X}[V2X]{vehicle-to-everything}
    \acro{FL}[FL]{Federated Learning}
    \acro{CL}[CL]{Centralized Learning}
    \acro{ML}[ML]{Machine Learning}
    \acro{SGD}[SGD]{Stochastic Gradient Descent}
    \acro{FedAvg}[FedAvg]{Federated Averaging}
    \acro{IoT}[IoT]{Internet of Things}
    \acro{CPS}[CPS]{Cyber-Physical Systems}
    \acro{OBD}[OBD]{On-Doard Diagnostics}
    \acro{DoS}[DoS]{Denial of Service}
    \acro{FedProx}[FedProx]{Federated Proximal}
    \acro{iid}[i.i.d]{Independent and Identically Distributed}
    \acro{CIDS}[CIDS]{Clock-Based Intrusion Detection System}
    \acro{NIDS}[NIDS]{Network Intrusion Detection System}
    \acro{HIDS}[HIDS]{Host Intrusion Detection System}
    \acro{SIDS}[SIDS]{Signature-based Intrusion Detection System}
    \acro{AIDS}[AIDS]{Anomaly-based Intrusion Detection System}
    \acro{CNN}[CNN]{Convolutional Neural Network}
    \acro{RNN}[RNN]{Recurrent Neural Network}
    \acro{MQTT}[MQTT]{Message Queuing Telemetry Transport}
    \acro{FPR}[FPR]{False Positive Rate}
    \acro{qos}[QoS]{Quality of Service}
    \acro{DR}[DR]{Detection Rate}
\end{acronym}

\title{Evaluating the Impact of Privacy-Preserving Federated Learning on CAN Intrusion Detection}

\author{\IEEEauthorblockN{
Gabriele Digregorio,
Elisabetta Cainazzo, 
Stefano Longari,
Michele Carminati, and
Stefano Zanero \\
Politecnico di Milano, Milan, Italy \\
\{gabriele.digregorio, stefano.longari, michele.carminati, stefano.zanero\}@polimi.it\\
elisabetta.cainazzo@mail.polimi.it
}}

\maketitle
{\renewcommand{\thefootnote}{}\footnotetext{This paper has been accepted at the IEEE 99th Vehicular Technology Conference (VTC2024-Spring), 2024.}}

\begin{abstract}
The challenges derived from the data-intensive nature of machine learning in conjunction with technologies that enable novel paradigms such as V2X and the potential offered by 5G communication, allow and justify the deployment of \ac{FL} solutions in the vehicular intrusion detection domain.
In this paper, we investigate the effects of integrating \ac{FL} strategies into the machine learning-based intrusion detection process for on-board vehicular networks. Accordingly, we propose a \ac{FL} implementation of a state-of-the-art \ac{IDS} for Controller Area Network (CAN), based on LSTM autoencoders. We thoroughly evaluate its detection efficiency and communication overhead, comparing it to a centralized version of the same algorithm, thereby presenting it as a feasible solution.
\end{abstract}

\begin{IEEEkeywords}
Controller Area Network, Federated Learning, Intrusion Detection
\end{IEEEkeywords}

\section{Introduction}
The shift of the automotive industry towards a more connected and autonomous landscape, while offering increased functionality and convenience, also makes automotive systems more susceptible to cyber-attacks. Amongst the security measures against such threats, \acfp{IDS} for automotive on-board networks are becoming a popular tool for identifying and addressing unusual activities. \ac{ML} enhances the performance of such \acp{IDS} by processing and learning from large datasets, but its data-heavy approach poses challenges in automotive contexts.
Creating effective \acp{IDS} models demands extensive data reflecting diverse driving conditions and high computational power for training and deployment, often exceeding in-vehicle system capabilities.
Hence, up to now, training algorithms for vehicular contexts have mostly relied on \ac{CL}, which collects and stores data from multiple vehicles in a centralized location, where the training process takes place. 

Emerging technologies like V2X, edge computing, and advancements in data communication with 5G and upcoming 6G, have largely addressed the challenges of requiring numerous vehicles to transmit substantial amounts of data to a centralized server. However, sending raw vehicle data to a central node raises privacy issues, as this data can contain personal and sensitive information that ideally should not be shared with a central server.
\acf{FL} addresses these concerns allowing \acp{IDS} to benefit from the collective knowledge of the entire system without compromising individual privacy, as the raw information does not need to be shared.

Implementing federated versions of effective \ac{CAN} \acp{IDS} could be the key to their feasibility in real-world scenarios. This approach addresses the challenges of limited dataset diversity and the privacy issues related to transmitting vast amounts of \ac{CAN} data from vehicles to a centralized training location.
However, implementing a \ac{ML} algorithm in a federated manner introduces its own challenges, particularly in integrating data from various entities.

This paper proposes and assesses the viability of using \ac{FL} algorithms for intrusion detection within the automotive sector. We have developed a federated version of a state-of-the-art \ac{ML}-based \ac{IDS} for CAN, CANdito~\cite{longari2023candito}, and examined its effectiveness by comparing it with a centralized version of the same algorithm.

Our experimental evaluation focused on the tradeoffs in terms of detection capabilities and communication overhead in \ac{FL} approaches. We found that the volume of data each participant needs to transmit in a federated setup is greater than in a corresponding centralized model. However, this increased data requirement is reasonable, thanks to the potential offered by advancements in communication technologies. While the detection capabilities of the federated model are slightly lower compared to the centralized model, they still demonstrate robust performance. This slight reduction in detection effectiveness is a reasonable cost to pay for the substantial privacy benefits that \ac{FL} offers, addressing one of the key challenges in modern data-driven applications.

By using a \ac{LSTM} autoencoder-based \ac{ML} algorithm, we address limitations in current literature, namely the limited use of \ac{RNN} in \ac{FL} for CAN bus anomaly detection. Additionally, we explore the relatively new area of applying federated algorithms to \ac{LSTM} autoencoders.

In short, our contributions are the following:
\begin{itemize}
    \item We propose a federated approach for intrusion detection in on-board vehicular networks based on CANdito~\cite{longari2023candito}, a state-of-the-art LSTM autoencoder-based \ac{IDS} for CAN.
    \item We extensively evaluate the performances of our approach against a centralized version of the same \ac{IDS}, demonstrating comparable detection capabilities of the federated version in relation to its centralized counterpart.
    \item We assess the communication overhead of MQTT over 5G during the training rounds of the federated implementation.
\end{itemize}

\section{Background and Motivation}
\label{ch:chapter_two}%

\subsection{CAN security}
\label{sec:Control Area Network}

The \ac{CAN} protocol~\cite{specification1991bosch} is the industry standard for intra-vehicle communication. Its widespread adoption can be attributed to several key features: low cost, high resilience to interference, robust error detection, and the capability to handle numerous short messages in a multi-master system, making it well-suited for real-time applications.
A drawback of the age and simplicity of the \ac{CAN} protocol is that it lacks embedded security measures. It uses broadcast communications without cryptography protection, and identifiers are not authenticated, which leaves these networks open to packet injection, deletion, and modification. Attacks against CAN include: 
\begin{inparaenum}
    \item \ac{DoS} attacks that flood the bus with a vast amount of high-priority messages, resulting in communication disruption for legitimate \acp{ECU}, impacting vehicle functionalities; 
    \item Injection attacks that involve inserting messages into the \ac{CAN} bus introducing unauthorized commands or data that could alter the vehicle's physical behavior. The stealthiness of attacks depends on tactics employed such as replaying payloads from previously observed packets (replay attacks), gradually adjusting sensor and actuator values to avoid abrupt changes (seamless change attacks), or modifying the timing of packet delivery;
    \item Drop attacks that delete legitimate packets, disrupting communication flow and potentially resulting in the loss of critical data essential for the correct functioning of the vehicle;
    \item Masquerade attacks that usually combine drop and injection tactics to mimic behaviors of legitimate ECU. This method inserts unauthorized commands or data onto the bus while maintaining packet arrival frequency, evading detection.    
\end{inparaenum}


\subsection{CAN Intrusion Detection}
\label{subsec:IDS types}
Intrusion detection for vehicular systems can be roughly divided into hardware-, specification-, flow-, and payload-based detection. 
\textbf{Hardware-based detection}~\cite{DBLP:journals/spl/MurvayG14} fingerprints the \acp{ECU} physical characteristics. Since only a specific \ac{ECU} is allowed to send a given ID, a mismatch between the packet ID and the \ac{ECU} fingerprint may indicate a data injection attack by an attacker that is spoofing a different \ac{ECU}'s ID. While effective in detecting injection attacks, it usually requires extra hardware to generate the fingerprint and may not recognize misbehavior if the attacker has direct control of targeted \ac{ECU}. 
\textbf{Specification-based detection} focuses on detecting misbehavior in the use of the \ac{CAN} protocol, e.g. by monitoring the network to detect ID conflicts~\cite{dagan2016parrot} or detecting if an \ac{ECU} is disconnected from the \ac{CAN} network~\cite{DBLP:conf/ccs/LongariPCZ19}. While effective, it usually requires to be installed in all \acp{ECU} and additional hardware. 
\textbf{Flow-based detection} analyzes packet flow on the network, focusing on the arrival frequency of packets with identical IDs~\cite{DBLP:conf/wcicss/TaylorJL15,DBLP:conf/codaspy/YoungOBZ19} or the sequence of ID appearances on the bus~\cite{DBLP:journals/tits/IslamRYM22}. These methods suit the \ac{CAN} protocol due to its message flow regularity and predictability. However, they might not detect advanced masquerade attacks, where the ID periodicity is maintained but the payload is altered.
\textbf{Payload-based detection} scrutinizes packet contents, studying temporal relationships within packets~\cite{longari2021cannolo,longari2023candito}, examining correlations between packets with different IDs in the same timeframe~\cite{DBLP:journals/vcomm/SongWK20}, or employing a mix of these methods~\cite{DBLP:journals/access/HanselmannSDU20}. 
They can handle complex patterns not identifiable through simple rules but face practical limitations due to high computation and training demands.
For a comprehensive overview of existing solutions for CAN intrusion detection, we refer the reader to~\cite{DBLP:journals/comsur/LampeM23}. 

A vast subset of works in this area focuses on employing \ac{ML} techniques to distinguish between normal and anomalous patterns. 
Deep learning techniques, especially time-series analysis with RNNs and LSTM autoencoders, have proven to be effective recently~\cite{DBLP:journals/comsur/LampeM23, longari2023candito}.

\subsection{Collaborative Learning: Centralized vs Federated}

Training effective \acp{IDS} requires substantial data and computational capabilities difficult to fulfill by single vehicles. Hence, cooperation between vehicles is required. This cooperation can follow the different principles of \acf{CL} and \acf{FL}, which differ in terms of use cases, system requirements, communication costs, privacy concerns, and the level of cooperation.

\mypar{Centralized Learning} The \ac{CL} strategy involves a central entity, like a roadside infrastructure or a remote server, that collects, processes, and coordinates vehicle information. Each vehicle transmits its raw data to the central entity and then the central entity performs all the necessary computation and decision-making processes. However, this approach raises concerns about vehicle data privacy.

\mypar{Federated Learning} The \ac{FL} approach addresses some drawbacks of \ac{CL}. In \ac{FL} applied to the automotive sector, each vehicle in the network has its processing capabilities and shares summarized or aggregated data with the remote server, instead of the entire dataset. This ensures that sensitive information remains localized within each vehicle, thus reducing the risk of privacy breaches. A \ac{FL} strategy is usually composed of several communication rounds. At each round, a group of vehicles independently trains a \ac{ML} model using their local data. Then, rather than sharing the entire model or data, each vehicle exchanges only model updates with a central server. The central server strategically combines the received model updates to create a global model, which is then shared back with each vehicle. This process is iteratively repeated.
\section{Approach}
\label{sec:approach}
\begin{figure*}
    \centering
    \includegraphics[width = .99\textwidth]{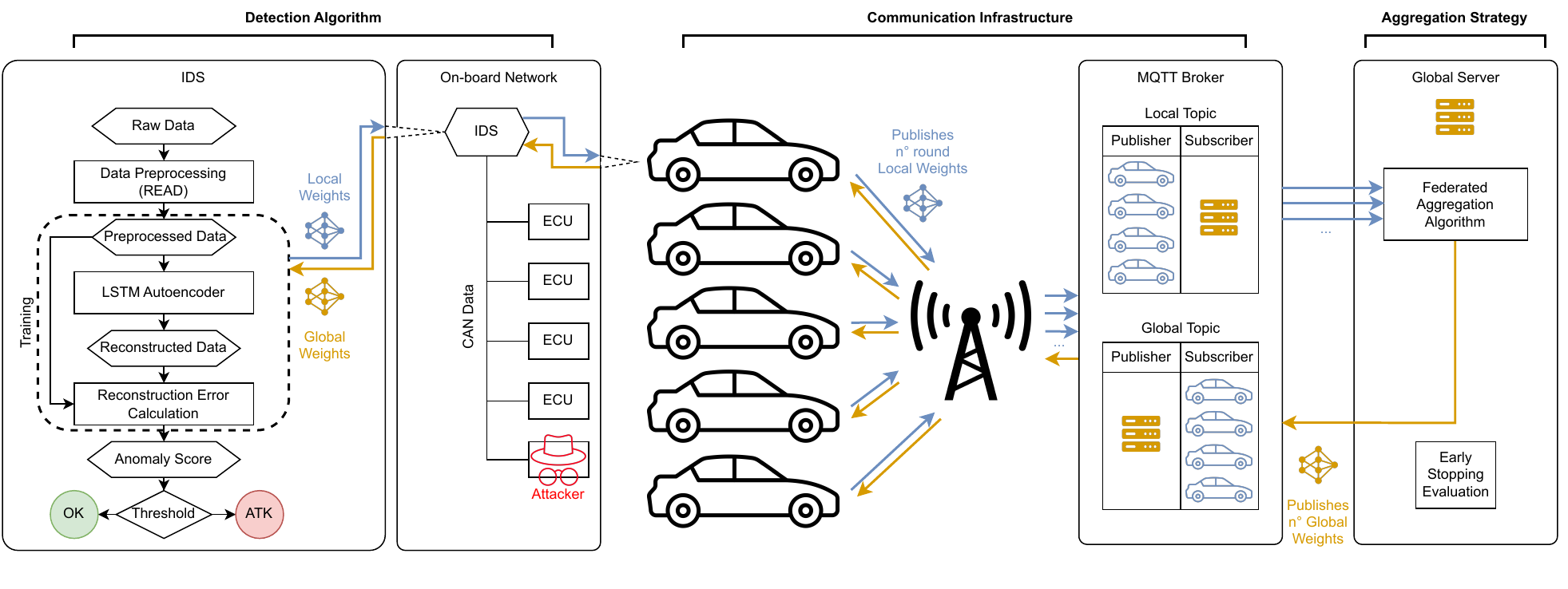}
    \caption{Overview of our system. From the left, the intrusion detection process shows the detection steps of CANdito~\cite{longari2023candito}, which - in the training phase - feed the local weights through the 5G communication of the vehicle to the MQTT broker, which provides them to the Global Server. The server computes the global weights for the round, decides whether early stopping is necessary, and provides the weights to the vehicles.}
    \label{fig:overview}
\end{figure*}

Our approach, as illustrated in Figure~\ref{fig:overview}, comprises three primary components: a detection algorithm, a communication infrastructure, and a federated aggregation strategy. 
It incorporates an \ac{LSTM} autoencoder-based \ac{IDS} for \ac{CAN}, deployed across a vehicle fleet and trained using \ac{FL}.
In each round of federated training, detection systems transmit their local model weights to a designated \ac{MQTT} topic via their vehicle's 5G connection. The Global Server, which subscribes to this topic, collects this data. It computes the global weights for the round using a federated aggregation algorithm, such as FedAvg~\cite{DBLP:conf/aistats/McMahanMRHA17} or FedProx~\cite{DBLP:conf/mlsys/LiSZSTS20}. Subsequently, the Global Server distributes these global weights back to the vehicles by posting them to a different \ac{MQTT} topic, to which all vehicles are subscribed. Additionally, in each round, the Global Server determines whether to continue or halt the training process based on early stopping criteria.

\subsection{Detection Algorithm}
We identified CANdito~\cite{longari2023candito} as the best candidate for our approach, given the results demonstrated both in detection performances and in computation time, which--especially combined--outperform current state-of-the-art detection systems for CAN.
CANdito relies on \ac{LSTM}-autoencoders, which has shown to be an effective detection method for CAN~\cite{DBLP:journals/comsur/LampeM23}, and is based on the assumption that a vehicle's behavior can be seen as a sequence of finite events where each event depends on the previous ones. At training time, the autoencoder analyzes legitimate data streams from the CAN bus and builds a latent representation of the CAN traffic data without requiring knowledge of data semantics. At runtime, the autoencoder attempts to reconstruct the CAN traffic in the sequence. It then computes the reconstruction error as the difference between the forecasted packets and the actual packets in the sequence. If the reconstruction error is above a certain threshold, attacks or anomalies are detected. 

\mypar{Preprocessing and data acquisition}
Feeding the raw CAN stream to CANdito requires specifying the location and type of signals within the payload for each CAN ID. This step has to be coherent for all vehicles participating in the \ac{FL} process. 
Following the preprocessing steps of the previous work~\cite{longari2023candito}, we employed READ~\cite{DBLP:journals/tifs/MarchettiS19} to identify signals and their categories by analyzing the frequency of bit changes in CAN payloads.
While this process would not be necessary for a carmaker, which has access to the signal definitions, it has already been used multiple times as an effective alternative~\cite{DBLP:journals/compsec/NicheliniPLCZ23,longari2021cannolo,longari2023candito}. Once this segmentation and categorization are known, the CANdito algorithm requires to be fed time series of 40 sequential packets with the same ID to detect an intrusion.

\mypar{Threshold in the federated algorithm}
We assessed different configurations for calculating the optimal model threshold for the reconstruction error, involving different levels of decentralization. In these setups, each vehicle computes its optimal model threshold using a small, local dataset not previously used, after receiving the final model from the Global Server. These individual thresholds are then sent to the Global Server, which aggregates them to determine the final threshold. However, this method resulted in significant performance degradation due to the model's high sensitivity to threshold accuracy. We also found that the threshold computation is a relatively lightweight operation compared to the whole training process. Therefore, in our final implementation, the threshold is computed by a single vehicle, which then shares the computed value with others. This designated vehicle could be one that has already participated in the training process, possessing sufficient data and resources, or it could be an additional vehicle specifically tasked with threshold computation, rather than participating in training.

\mypar{Federated Early Stopping}
To mitigate the risk of overfitting and optimize the number of training rounds, we implemented a decentralized early stopping strategy. After each global model update, every vehicle evaluates the model's performance using the Mean Squared Error (MSE) loss on a small local validation set. The vehicles then share their individual validation losses to the Global Server, which computes an average to obtain a global loss. This method is similar to centralized early stopping: training ceases if the global loss does not show improvement for a number of consecutive rounds, as specified by the ‘patience’ parameter. The minimum magnitude of improvement between rounds is quantified by the $\delta$ parameter.

\subsection{Communication Infrastructure}
For update sharing, we implemented a publish-subscribe system via the MQTT 5.0 over TCP protocol. We enabled TLS encryption and authentication to ensure update confidentiality and prevent unauthorized vehicles from submitting updates.
We chose Eclipse Mosquitto~\cite{mosquitto_website}, an open-source message broker, in its latest version as of January 2024. To ensure accurate update delivery, we set the \ac{qos} level to ‘exactly once delivery’ (\ac{qos} 2)~\cite{mqtt2019}.
The infrastructure we designed includes two distinct topics. 
The first topic, or the 'local' topic, is used by each vehicle to publish its local updates, primarily consisting of weight updates resulting from local training.
The second topic, or the 'global' topic, is designated for the Global Server, which publishes global updates after aggregating and averaging vehicle updates.
This setup ensures an efficient and organized exchange of information between the vehicles and the Global Server.

\subsection{Aggregation Strategy}
We focused on \ac{FedAvg}~\cite{DBLP:conf/aistats/McMahanMRHA17} and \ac{FedProx}~\cite{DBLP:conf/mlsys/LiSZSTS20}, two \ac{FL} algorithms, which differ primarily in the local objective function employed.

\mypar{FedAvg~\cite{DBLP:conf/aistats/McMahanMRHA17}} Each node in the network downloads an initial global model and locally improves it by training on a local dataset over a specified number of epochs. After this, each node sends its model updates to a global server. The server aggregates these updates to produce a new global model, which is then redistributed to the nodes for further local training epochs.
This cycle of local training and aggregation continues until the global model attains the targeted accuracy or meets other predefined criteria.
The two principal parameters of FedAvg are the \textit{number of epochs} $E$ and the \textit{number of communication rounds} $R$. The number of epochs refers to the iterations of training each node performs before transmitting its weights to the global server in a round. 
The number of communication rounds indicates how often the nodes interact with the global server sending their locally trained model weights.
FedAvg is efficient and helps address privacy concerns since raw data remains within its original node. Nonetheless, it demands careful management of aspects like network bandwidth, heterogeneous data distributions, and the handling of non-i.i.d. (independent and identically distributed) data across nodes.

\mypar{FedProx~\cite{DBLP:conf/mlsys/LiSZSTS20}} 
It extends the principles of FedAvg introducing a \textit{proximal term} $\mu$ to the local objective function, which ensures local updates align closely with the global model. This enhances stability in non-i.i.d. environments with high heterogeneity.
On the other hand, FedProx adds complexity due to the need for careful tuning of the proximal term, balancing between local and global model accuracy.
\section{Experimental Evaluation}
Previous studies~\cite{DBLP:conf/aistats/McMahanMRHA17,DBLP:journals/corr/abs-1812-01097} have highlighted the challenges associated with the selection of hyperparameters for FedAvg and FedProx, noting that improper choices can lead to divergence and suboptimal outcomes. 
Hence, our experimental evaluation aims to optimize local training epochs per round ($E$) for FedAvg and FedProx, optimize the proximal term ($\mu$) for FedProx, evaluate decentralization's effect on model convergence across different vehicle counts ($V$), and compare the federated CANdito's efficiency and communication costs to its centralized version, trained on the same dataset.

\subsection{Dataset Overview}
For our experiments, we used the ReCAN C-1 dataset~\cite{recan}, a real-world CAN traffic dataset recorded in an Alfa Romeo Giulia Veloce during both city and highway driving. The dataset is divided into 9 driving sessions, totaling 25,082,275 packets. For our experiments, we used data from driving session numbers 1, 2, 6, 8, and 9 for the training stage, data from driving session number 5 for the validation stage, and data from driving session number 7 for the test stage. 
The presence of data from different driving sessions makes the setting non-i.i.d., potentially influencing the convergence capabilities of the employed federated algorithms. On the other hand, this setting makes the experiments more representative of real-world scenarios and hence more relevant. 
We evaluated the performance of our approach on 13 selected CAN IDs, chosen based on their use in CANdito~\cite{longari2023candito} and CANova~\cite{DBLP:journals/compsec/NicheliniPLCZ23}.

\mypar{Attack generation}
\label{sec:attack_generation}
We injected attacks into both the validation and test datasets using the CANtack tool, already proposed and used in the literature~\cite{longari2023candito,DBLP:journals/compsec/NicheliniPLCZ23} and available online\footnote{https://bitbucket.org/necst/attack\_tool\_code}. The tool allows for the simulation and injection of a wide range of attacks in the ReCAN datasets. Our methodology aims to challenge the IDS's effectiveness against both basic and sophisticated attackers. To this end, we created injection attacks by introducing a sequence of 25 packets into the datasets. For masquerade attacks, which simulate an advanced adversary taking control of a CAN node and transmitting on its behalf, we altered the payload of existing packets.
The attack strategies include signal data fuzzing, which avoids obvious detection by not changing bits that are static in authentic payloads, executing replay attacks, and avoiding unrealistic changes subtly shifting signal values to their extreme limits from the last genuine packet's value. Additionally, we conducted drop attacks by removing a sequence of 25 consecutive packets with the targeted ID from the dataset.

\begin{table}[t]
\centering
\caption{Dataset size (in number of packets) for each vehicle for different levels of decentralization. $V$ represents the number of vehicles participating in the learning process.}
\label{tab:dataset_size}
\begin{tabular}{c|c|c|c|c|c}
\toprule
ID & CL & V = 5 & V = 10 & V = 20 & V = 50 \\

\midrule
0DE & 722000 & 144400 & 72200 & 36100 & 14440 \\
0EE & 724830 & 144966 & 72483 & 36241 & 14496 \\
0FB & 721331 & 144266 & 72133 & 36066 & 14426 \\
0FC & 721330 & 144266 & 72133 & 36066 & 14426 \\
0FE & 724830 & 144966 & 72483 & 36241 & 14496 \\
0FF & 721333 & 144266 & 72133 & 36066 & 14426 \\
1F7 & 363168 & 72633 & 36316 & 18158 & 7263 \\
1FB & 360922 & 72184 & 36092 & 18046 & 7218 \\
11C & 724827 & 144965 & 72482 & 36241 & 14496 \\
100 & 721333 & 144266 & 72133 & 36066 & 14426 \\
104 & 726338 & 145267 & 72633 & 36316 & 14526 \\
116 & 724828 & 144965 & 72482 & 36241 & 14496 \\
192 & 724350 & 144870 & 72435 & 36217 & 14487 \\

\bottomrule
\end{tabular}
\end{table}

\subsection{Experimental Results}
\mypar{Federated Algorithm Convergence}
\label{sec:federated_alg_vs}
We validated the performance of both FedAvg and FedProx using the validation dataset containing the injected attacks. 
For both FedAvg and FedProx, we distributed the training data across 5, 10, 20, and 50 vehicles ($V$) and conducted training with local epochs ($E$) set at 1, 3, and 5. For FedProx, we tested four proximal term ($\mu$) values: 1, 0.1, 0.01, and 0.001.
Table \ref{tab:dataset_size} displays the training data size for the \ac{CL} case and for different values of $V$.

\begin{figure}[t]
    \centering
    \includegraphics[width =\columnwidth]{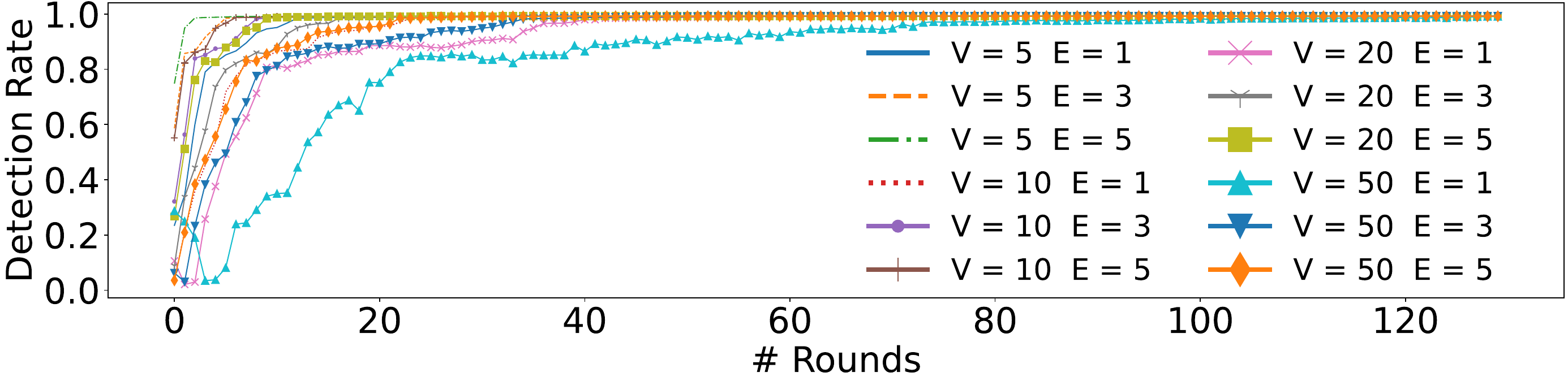}
    \caption{FedAvg \acl{DR} convergence results for different levels of decentralization. $V$ represents the number of vehicles participating in the learning process and $E$ denotes the number of local epochs that each vehicle trains during each round.  }
    \label{fig:fedavg_plot}
\end{figure}

For FedAvg, we trained each configuration over 200 communication rounds ($R$). Figure \ref{fig:fedavg_plot} depicts the average performance in terms of \ac{DR}, varying the values of $E$ and $V$, specifically focusing on CAN ID 192. To more effectively highlight the differences in convergence speed among the various settings, the figure only displays the first 130 rounds. Beyond 130 rounds, the performance trend becomes almost stationary around the same value for all the model settings. Increasing the value of $E$ marginally accelerates the convergence, whereas increasing the number of vehicles $V$ results in a delayed convergence point in terms of rounds $R$. This outcome aligns with expectations, as higher values of $E$ and lower values of $V$ more closely resemble a \ac{CL} setup. For the FedProx algorithm, we extended the training to 500 communication rounds for each configuration. This decision is based on a slower convergence speed observed during the validation process (omitted due to space constraints), compared to FedAvg. While the results across different settings of $E$ and $V$ are in line with those observed for FedAvg, the introduction of the proximal term $\mu$ in FedProx does not result in performance improvements. Conversely, increasing the value of $\mu$ appears to further slow down the convergence process.

\begin{figure*}[t]
    \centering
    \includegraphics[width =0.85\textwidth]{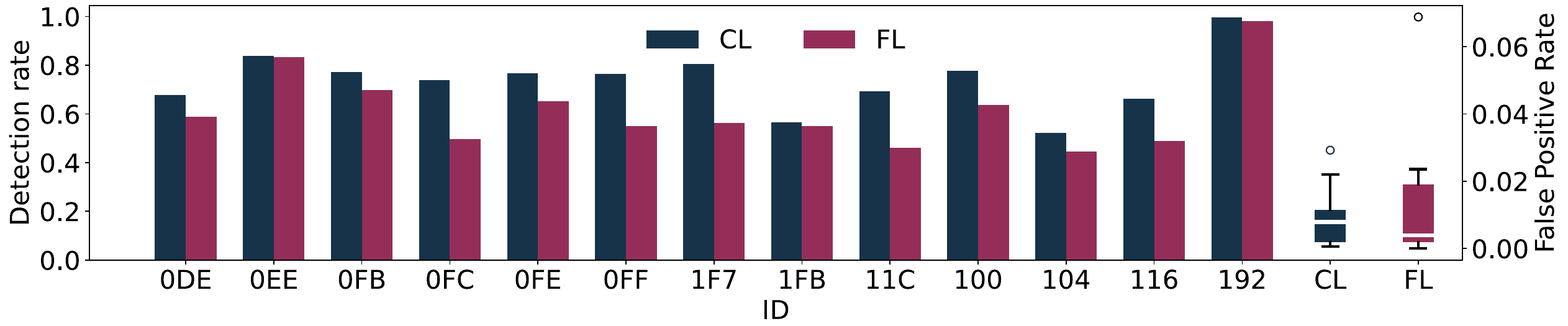}
    \caption{Average \acl{DR} on all attacks for each CAN ID and average \acl{FPR} on all attacks and all CAN IDs for centralized (CL) and 50-vehicle federated (FL) models, computed across attacks on the test set.}
    \label{fig:cl_vs_fl}
\end{figure*}

\mypar{Performance Overhead}
\label{subsec:cl_vs_fl}
This test aims to assess the federated CANdito \ac{IDS}'s effectiveness in detecting \ac{CAN} bus attacks, focusing on performance trade-offs caused by decentralization. We compare \ac{DR} and \ac{FPR} of the federated CANdito against its centralized version, trained on the same dataset.
We focused on the FedAvg algorithm due to its superior convergence performance, which reduces the number of rounds needed for the federated process while maintaining or improving performance compared to FedProx. To better evaluate decentralization's effect on CANdito's attack detection capabilities, we selected the most decentralized scenario from Section \ref{sec:federated_alg_vs}, involving 50 vehicles ($V=50$) and one local training epoch per vehicle in each round ($E=1$).
We applied federated early stopping with a patience setting of 10 rounds and a dynamic minimum $\delta$ of 3\% of the loss value. Our validation tests (omitted for brevity) indicated that different values for the patience and $\delta$ parameters result in either poorer performance or stalling during the training process.

Figure \ref{fig:cl_vs_fl} shows the comparative results between the centralized and federated versions of the CANdito IDS.
Overall, the federated model exhibits good detection capabilities, though it does not reach the performance of its centralized counterpart. This result aligns with our expectations, as the difference in performance can be ascribed to the approximations introduced in the learning process by the federated settings. 
The boxplot depicting the \ac{FPR} reveals for the federated CANdito model a marginally greater spread of \ac{FPR} values around 0.1 compared to the centralized model. Both models exhibit \ac{FPR} outliers, but the federated version tends to have outliers with slightly higher absolute values. Conversely, the federated model demonstrates a lower median \ac{FPR}, suggesting that it maintains low \ac{FPR} across a broader range of IDs than its centralized counterpart.

\mypar{Communication Overhead}
We consider the communication overhead caused by the decentralization of CANdito \ac{IDS}. Each test was conducted over a 5G network in a crowded university area of Milan for a few hours. To simulate a realistic scenario where the server is rarely close to the vehicles, the MQTT broker was placed on a remote server in London. We assessed the latency involved in a vehicle publishing a new model update to the ‘local’ topic and in receiving a global update from the ‘global’ topic by the Global Server. In this discussion, we do not account for the communication resources used to transmit the local validation loss to the Global Server for federated early stopping. This decision is based on the negligible size of the payload associated with these loss values when compared to the size of the model updates.

Our tests showed model updates to be highly homogeneous, with an average update size of 372,893 bytes and a standard deviation of 8 bytes. The largest and smallest payloads were 372,898 bytes and 372,862 bytes, respectively. 
We further evaluated latency by simulating 10,000 local and 7,500 global updates, with the average latency detailed in Table \ref{tab:subscribe_publish_results}. From these findings, Table \ref{tab:data_overhead} estimates the average communication overhead from the perspective of a single vehicle, using the most decentralized settings in Section \ref{subsec:cl_vs_fl}.
For each model trained on an ID, we consider the number of rounds $R$ required before the federated early stopping mechanism terminates the learning process. The metrics evaluated include the download time for global model updates (DL Time), the time needed to publish local updates (UL Time), the total raw data size downloaded for global updates (DL Data), and the total raw data size uploaded for local updates (UL Data). Additionally, the metric $\delta Data$ measures the difference in the amount of raw data exchanged (both download and upload) by each vehicle to complete the \ac{FL} process, compared to a \ac{CL} scenario where vehicles only upload their CAN raw data to a remote server and download the final model.
As shown by Table \ref{tab:data_overhead}, the \ac{FL} process incurs a significant overhead in terms of the amount of raw data exchanged compared to \ac{CL}. This increase is due to several factors: the number of rounds needed for convergence, the size of the model updates, the nature of CAN data that makes them lightweight w.r.t. other types of data used in \ac{ML}, and the relatively small amount of data possessed by each vehicle in a highly federated setting involving 50 vehicles.

\begin{table}[t]
\centering
\caption{Latency results for MQTT publish and receive test.}
\begin{tabular}{c|c|c|c|c|c}
\toprule
            & Average             & Std                & Median             & Min                & Max               \\ 

            \midrule
Subscriber   & 0.180s            & 0.040s           & 0.169s           & 0.066s           & 0.726s          \\ 
Publisher     & 0.411s            & 0.149s           & 0.365s           & 0.229s           & 2.990s          \\

\bottomrule
\end{tabular}
\label{tab:subscribe_publish_results}
\end{table}

\begin{table}[tb]
\centering
\caption{Communication Overhead of the Federated Learning process.}
\begin{tabular}{c|c|c|c|c|c|c}
\toprule
ID  & R  & DL Time & UL Time & DL Data & UL Data            & $\delta Data$               \\
& & (s) & (s) & (MiB) & (MiB) & (MiB) \\

\midrule
0DE & 104 & 18.72 & 42.74 & 37.34 & 36.98 & +72.56 \\ 
0EE & 42 & 7.56 & 17.26 & 15.29 & 14.94 & +28.17 \\ 
0FB & 79 & 14.22 & 32.47 & 28.45 & 28.09 & +54.49 \\
0FC & 83 & 14.94 & 34.11 & 29.87 & 29.52 & +57.34 \\ 
0FE & 183 & 32.94 & 75.21 & 65.43 & 65.08 & +128.45 \\
0FF & 23 & 4.14 & 9.45 & 8.53 & 8.18 & +14.66 \\ 
1F7 & 117 & 21.06 & 48.09 & 41.96 & 41.61 & +82.36 \\ 
1FB & 135 & 24.3 & 55.48 & 48.36 & 48.01 & +95.17 \\
11C & 58 & 10.44 & 23.84 & 20.98 & 20.63 & +39.55 \\ 
100 & 101 & 18.18 & 41.51 & 36.27 & 35.92 & +70.14 \\ 
104 & 175 & 31.5 & 71.92 & 62.59 & 62.23 & +122.76 \\ 
116 & 37 & 6.66 & 15.21 & 13.51 & 13.16 & +24.61 \\ 
192 & 118 & 21.24 & 48.5 & 42.32 & 41.96 & +82.66 \\ 

\bottomrule

\end{tabular}
\label{tab:data_overhead}
\end{table}

\subsection{Discussion}
The results presented for the centralized version of CANdito are based on the assumption that all data, which in the federated setting are distributed among vehicles and always kept local, are instead aggregated by a central server. While we demonstrated that a decentralized approach might be relatively disadvantageous in terms of communication overhead and detection capabilities, it also avoids sending potentially sensitive local CAN bus data to a central remote server. Given that underestimating privacy and security concerns is not an option in real-world applications, a centralized model becomes less viable, especially when extensive data are required for training a \ac{ML} model.

Our analysis indicates that in a federated setting, each contributor transmits more data than it would by simply sending its local dataset to a remote server and then receiving the final model directly. Nevertheless, this increase in data exchange is sustainable and aligns with advancements in data communication technologies like 5G and the forthcoming 6G. The slight reduction in detection capabilities of the federated CANdito compared to the centralized model, while still maintaining robust performance, represents a reasonable trade-off. This represents the cost of the significantly enhanced security and privacy that the federated approach offers, making it well-suited for real-world scenarios.

\section{Related Works}
\ac{FL} has emerged as a prominent distributed \ac{ML} paradigm that facilitates training models on decentralized data sources while preserving data privacy. McMahan et al. introduced the concept of \ac{FL} \cite{DBLP:conf/aistats/McMahanMRHA17}, enabling collaborative model training across mobile devices without transmitting raw data to a central server. Secure multi-party computation techniques are proposed by Bonawitz et al. \cite{DBLP:conf/ccs/BonawitzIKMMPRS17} to achieve secure aggregation of model updates while safeguarding individual data privacy. 
Optimization techniques leveraging stochastic gradient descent are explored to enhance the efficiency and convergence speed of \ac{FL}, as \ac{FedAvg}~\cite{DBLP:conf/aistats/McMahanMRHA17} and \ac{FedProx}~\cite{DBLP:conf/mlsys/LiSZSTS20}.
The application of \ac{FL} in healthcare \cite{info:doi/10.2196/medinform.7744,DBLP:journals/ijmi/BrisimiCMOPS18,DBLP:journals/jbi/HuangSQMDL19} and IoT \cite{DBLP:journals/iotm/ZhangGHZKA22,9475501} domains has demonstrated the feasibility of training models on distributed data while ensuring data privacy.
Scalability and communication efficiency are critical considerations as \ac{FL} scales to more participants and more complex models. Approaches such as hierarchical aggregation schemes \cite{DBLP:journals/corr/KonecnyMRR16} have been proposed to reduce communication overhead in large-scale federated settings. Additionally, model compression techniques, including knowledge distillation and quantization \cite{DBLP:conf/nips/AlistarhG0TV17}, are explored to mitigate the communication costs associated with \ac{FL}.
In the vehicular context, studies have examined the feasibility of \ac{FL} for \ac{ML}-based vehicular applications \cite{DBLP:conf/meditcom/ElbirSCGB22}, investigating object detection using image-based datasets as a case study. For in-vehicle networks, a practical privacy-preserving \ac{IDS} approach called ImageFed is proposed \cite{10132145}, utilizing federated \ac{CNN}. The robustness of ImageFed is evaluated in scenarios such as non-i.i.d. clients and limited training data availability during the FL process. Another work focuses on developing a \ac{CAN} bus anomaly detection system using Graph Neural Networks (GNN) \cite{DBLP:journals/tifs/ZhangZL23} to address the vulnerability of the \ac{CAN} bus to various attacks.
\section{Conclusion}
In this paper, we explored the use of \ac{FL} algorithms for intrusion detection in the automotive industry. Our work involved developing a federated version of the state-of-the-art \ac{ML}-based IDS for CAN known as CANdito and evaluating its performance against a centralized version of the same algorithm. The results of our experiments, which focus on detection capabilities and communication overhead, suggest that FL could be a suitable approach in real-world scenarios where ignoring data privacy and security is not an option.
\section*{Acknowledgement}
The research leading to these results has been partially funded by  MICS (Made in Italy – Circular and Sustainable) Extended Partnership and received funding from Next-Generation EU (Italian PNRR – M4 C2, Invest 1.3 – D.D. 1551.11-10-2022, PE00000004), CUP MICS D43C22003120001 and the Italian Ministry of University and Research (MUR) under the PRIN 2022 PNRR framework (EU Contribution – Next Generation EU – M. 4,C. 2, I. 1.1), SHIELDED project, ID P2022ZWS82, CUP D53D23016240001.

\bibliographystyle{ieeetr}

\bibliography{bibliography.bib}

\end{document}